\documentclass[twocolumn,twocolappendix]{aastex631}

\usepackage{natbib}
\usepackage{amsmath}
\usepackage{amssymb}

\newcommand{\unit}[1]{\hfill\text{}\mathrm{#1}}

\begin{document}

  \title{Protostellar chimney flues: \\ are jets and outflows lifting submillimetre dust grains from discs into envelopes?}

\correspondingauthor{Luca Cacciapuoti}
\email{luca.cacciapuoti@eso.org}

\author{L. Cacciapuoti}
\affiliation{European Southern Observatory, Karl-Schwarzschild-Stra{\ss}e 2, 85748 Garching bei Munchen, Germany}
\affiliation{Fakultat f\'{u}r Physik, Ludwig-Maximilians-Universit\"{a}t M\"{u}nchen, Scheinerstra{\ss}e 1, 81679 M\"{u}nchen, Germany}
\affiliation{INAF, Osservatorio Astrofisico di Arcetri, Largo E. Fermi 5, 50125, Firenze, Italy}

\author{L. Testi}
\affiliation{Dipartimento di Fisica e Astronomia "Augusto Righi" Viale Berti Pichat 6/2, Bologna}
\affiliation{INAF, Osservatorio Astrofisico di Arcetri, Largo E. Fermi 5, 50125, Firenze, Italy}


\author{L. Podio}
\affiliation{INAF, Osservatorio Astrofisico di Arcetri, Largo E. Fermi 5, 50125, Firenze, Italy}

\author{C. Codella}
\affiliation{INAF, Osservatorio Astrofisico di Arcetri, Largo E. Fermi 5, 50125, Firenze, Italy}
\affiliation{Univ. Grenoble Alpes, CNRS, IPAG, 38000 Grenoble, France}

\author{A.J. Maury}
\affiliation{Universit\'{e} Paris-Saclay, Universit\'{e} Paris Cité, CEA, CNRS, AIM, 91191, Gif-sur-Yvette, France}

\author{M. De Simone}
\affiliation{European Southern Observatory, Karl-Schwarzschild-Stra{\ss}e 2, 85748 Garching bei Munchen, Germany}
\affiliation{INAF, Osservatorio Astrofisico di Arcetri, Largo E. Fermi 5, 50125, Firenze, Italy}

\author{P. Hennebelle}
\affiliation{Universit\'{e} Paris-Saclay, Universit\'{e} Paris Cité, CEA, CNRS, AIM, 91191, Gif-sur-Yvette, France}

\author{U. Lebreuilly}
\affiliation{Universit\'{e} Paris-Saclay, Universit\'{e} Paris Cité, CEA, CNRS, AIM, 91191, Gif-sur-Yvette, France}

\author{R. S. Klessen}
\affiliation{Universität Heidelberg, Zentrum f\"{u}r Astronomie, Institut f\"{u}r Theoretische Astrophysik, Albert-Ueberle-Straße 2, 69120 Heidelberg, Germany}
\affiliation{Universit\"{a}t Heidelberg, Interdisziplin\"{a}res Zentrum f\"{u}r Wissenschaftliches Rechnen, INF 205, D-69120 Heidelberg, Germany}

\author{S. Molinari}
\affiliation{INAF-Istituto di Astrofisica e Planetologia Spaziali, Via del Fosso del Cavaliere 100, I-00133, Rome, Italy}

\begin{abstract}
Low dust opacity spectral indices ($\beta < 1$) measured in the inner envelopes of class 0/I young stellar objects (age $\sim 10^{4-5}$ yr) have been interpreted as the presence of (sub-)millimetre dust grains in these environments. The density conditions and the lifetimes of collapsing envelopes have proven unfavorable for the growth of solids up to millimetre sizes. As an alternative, magneto-hydrodynamical simulations suggest that protostellar jets and outflows might lift grains from circumstellar discs and diffuse them in the envelope. We reframe available data for the CALYPSO sample of Class 0/I sources and show tentative evidence for an anti-correlation between the value of $\beta_{1-3mm}$ measured in the inner envelope and the mass loss rate of their jets and outflows, supporting a connection between the two. We discuss the implications that dust transport from the disc to the inner envelope might have for several aspects of planet formation. Finally, we urge for more accurate measurements of both correlated quantities and extension of this work to larger samples, necessary to further test the transport scenario.
\end{abstract}

\section{Introduction}
\label{sec:intro}
The formation of terrestrial planets and of the rocky cores of giant planets is thought to happen in a core-accretion scenario, a process spanning ten orders of magnitude in size, where interstellar medium, sub-micron dust grains grow into km-sized objects. While dust growth has been long thought to take place exclusively in isolated, evolved protoplanetary discs revolving around class II young stellar objects (YSOs), recent observations indicate that dust growth up to millimeter sizes might start in collapsing protostellar envelopes, thus much earlier and further away from host stars than previously thought.
Observationally, the slope $\alpha$ of the spectral energy distribution (SED) across (sub-)millimetric wavelengths is a means to interpret interstellar dust properties and its size. Specifically, if (i) dust opacity scales as a power law ($\kappa \propto \nu^{\beta}$), (ii) the emission is optically thin, and (iii) the Rayleigh-Jeans (RJ) approximation holds, then $\beta = \alpha - 2$ (\citealt{Natta2007}, \citealt{Testi2014}). In turn, $\beta$ depends on dust properties, and strongly on the maximum grain size of the dust population. For the interstellar medium, typically $\beta \sim 1.7$ \citep{Weingartner2001}. In Class II objects, $\beta < 1$ suggests the presence of millimetre dust grains (e.g., \citealt{Testi2014}, \citealt{Tazzari2021}, \citealt{Macias2021}).
\begin{table*}
\scalebox{1.0}{
\begin{tabular}{c c c c c c c}
 \hline \hline
{Source} & $\dot M_{J}$ (B)$^a$  & $\dot M_{J}$ (R)$^a$ & $\dot M_{OF}$ (B)  & $\dot M_{OF}$ (R) & $\beta_{500 au}^b$ & M$_{env}^b$ \\
 & (10$^{-7}$ M$_{\odot}$/yr) & (10$^{-7}$ M$_{\odot}$/yr) & (10$^{-8}$ M$_{\odot}$/yr) & (10$^{-8}$ M$_{\odot}$/yr) &  & (M$_{\odot}$)\\
 \hline 
L1527      & $<0.2$ & $<0.2$       & $0.5$   & $0.5$   & 1.41 $\pm$ 0.16 & 1.2 \\
L1157      & 1.6      & 2.9        & 2.5     & 1.9     & 1.17 $\pm$ 0.18 & 3.0 \\
SVS13B     & $<0.1$   & $<0.1$     & $<0.02$ & $<0.02$ & 0.99 $\pm$ 0.16 & 2.8 \\
IRAS2A1    & $>6.7$   & $<0.1$     & 2.4     & 1.5     & 0.82 $\pm$ 0.17 & 7.9 \\
SerpS-MM18a& $>11.5$  & $>5.9$     & 12.7    & 5.1     & 0.74 $\pm$ 0.16 & 5.4  \\
SerpM-S68Na$^{(1)}$ & $>3.3$   & $<0.1$     & 0.12    & $<0.02$ & 0.66 $\pm$ 0.27 & 11.0 \\
IRAS4B1    & 1.2      & 2.3        & 3.6     & 1.0     & 0.62 $\pm$ 0.16 & 4.7 \\
IRAS4A1    &$\geq4.0$ & $\geq13.7$ & 12.4    & 11.7    & 0.54 $\pm$ 0.16 & 12.2 \\
L1448-C    & 9.7      & 13.8       & 7.3     & 10.1    & 0.41 $\pm$ 0.16 & 1.9 \\
\hline
\multicolumn{7}{l}{$^a$ from \citet{Podio2021} $^b$ from \citet{Galametz2019}}
\end{tabular}}
\caption{Dust opacity spectral indices at 500 au ($\beta_{500au}$), envelope masses ($M_{env}$) and jets and outflows mass loss rates ($\dot M_{J}$ and $\dot M_{OF}$, respectively) from the selected Class 0 sources from the CALYPSO sample \citep{Maury2019}. The $\dot M_{OF}$ are measured in this work and are lower limits as the outflow LV emission is likely optically thick. We report upper limits if no jet/outflow was detected in the CALYPSO data. (1) We do not include SerpM-S68N in the correlation because its SiO~($5-4$) transition prevents us from identifying high- and low-velocity ranges properly (we note, however, that the correlation is not affected by this point).}
\label{tab:1} 
\end{table*}

Several authors measured low $\beta$ values in the inner envelopes (a few $10^2$ au) of Class 0/I sources (\citealt{Kwon09}, \citealt{Miotello2014}, \citealt{Galametz2019}). Although $\beta$ also depends on grain composition and porosity, some of the observed values are too low ($\beta<0.5$) to be explained without considering 100 $\mu$m - 1 mm grains (e.g., \citealt{Kholer2015}, \citealt{Ysard2019}). However, simulations have so far predicted that dust coagulation would be ineffective at the low densities ($n\sim 10^{5-7}$ cm$^{-3}$) and short timescales (a few 10$^{5}$ yr) that characterize these environments (\citealt{Ormel2009}, \citealt{Lebreuilly2023}, \citealt{Silsbee2022}). It will be crucial for next simulations to test the effects of generally disregarded processes, like the dust back-reaction on the turbulence through gas-dust friction and dust-magnetic-field interaction \citep{2023arXiv230309883H}, to check whether growth remains a viable scenario.

Alternative or concomitant processes must be considered that could contribute to explain the observed low $\beta$. For example, \citet{Wong2016} first presented a simple analytical model to argue that millimetre dust from the disc could be entrained by protostellar outflows and transported to the envelope. \citet{Giacalone2019} also presented an analytical model for the entraiment of dust grains along magnetohydrodynamical (MHD) disc winds, and concluded that grains of $\sim$10 $\mu$m can be lifted by MHD winds and be transported outwards in the disc of T Tauri and Herbig Ae/Be objects. However, their model assumes typical evolved mass outflow rates of $\sim10^{-8}$ M$_{\odot}$/yr. Since the maximum grain sizes lifted in the envelopes depend linearly on this quantity, it can be much larger in young Class 0/I objects, for which the mass loss rates are orders of magnitude higher. These findings might have found recent confirmation thanks to exquisite JWST observations of the Tau 042021 edge-on disc, for which \citet{Duchene2023} reported an X-shaped feature in dust scattered whose spatial location is consistent with ALMA CO line emission tracing an outflow. Their observations suggest the entrainment of $\gtrsim 10 \mu$m grains even beyond 300 au. Finally, the findings of these models are supported by \citet{Tsukamoto2021} and \citet{Lebreuilly2023}, who arrived to consistent conclusions via three-dimensional MHD simulations. In particular, \citet{Tsukamoto2021} proposed the expression \textit{ash-fall}\footnote{Continuing \citet{Tsukamoto2021} nomenclature, we thus propose outflows as ``chimney flues'' in the title of this work.}, referring to the dust grains decoupling from the entrainment outflow and their subsequent fall back in the disc. 

Thus, as outflows represent in principle a means to transport submillimetre grains to envelopes, we here explore the unique CALYPSO sample \citep{Maury2019} to test whether a correlation holds between the observed power of jets and outflows in Class 0 protostars and the dust opacity index in their envelopes. 

\section{The sample}
\label{sec:sample}
\begin{figure*}
    \centering
    \includegraphics[width=14cm]{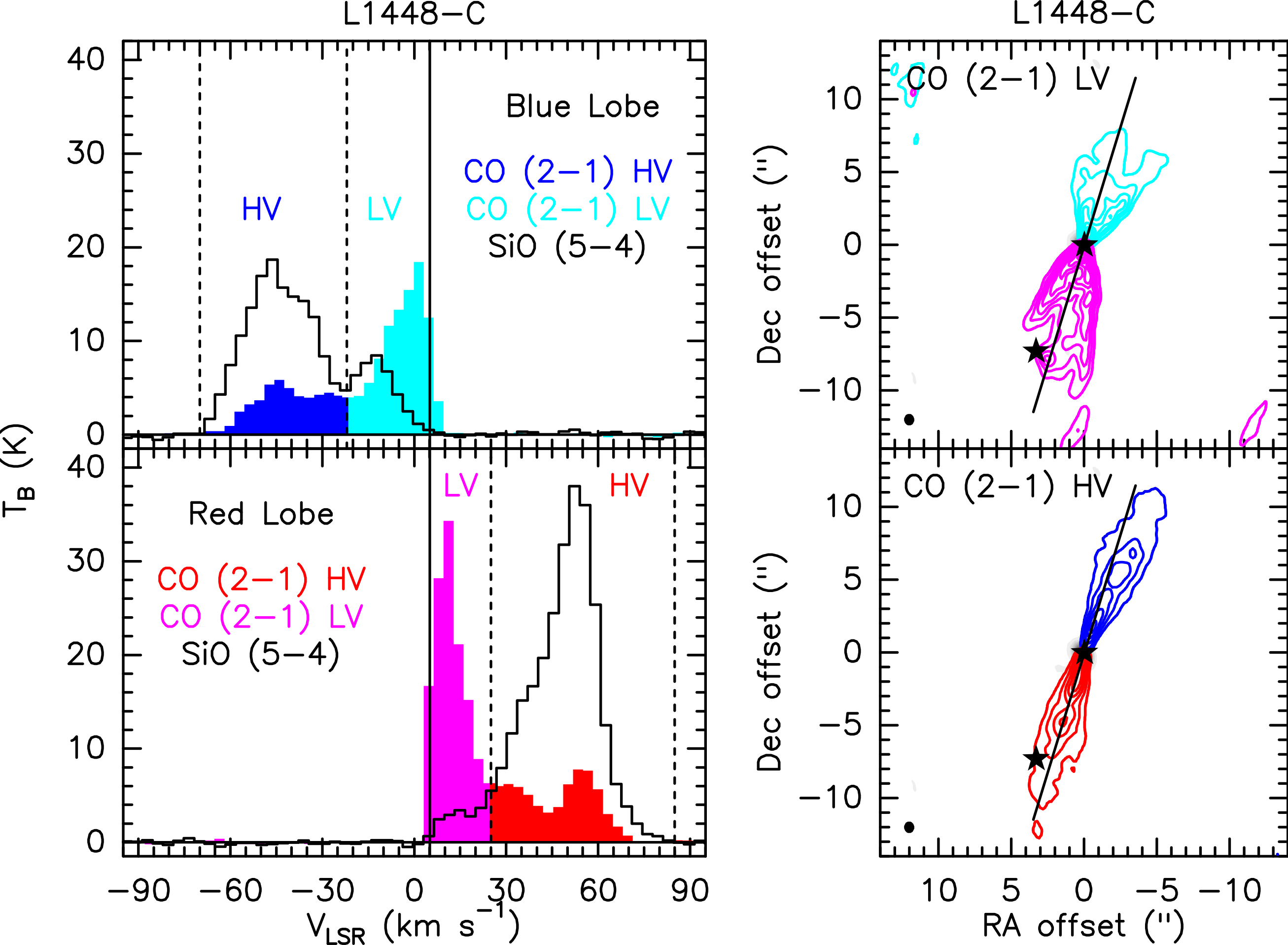}
    \caption{The L1448-C jet and the outflow, as imaged with PdBI in CO ($2-1$) and SiO ($5-4$). {\it Left panels:} CO (blue, cyan, magenta, and red) and SiO (black) spectra at the position of the blue-shifted and red-shifted SiO knots located closest to the driving source \citep[from ][]{Podio2021}. The horizontal and vertical solid lines indicate the baseline and systemic velocity, V$_{\rm sys} =+5.1$ km\,s$^{-1}$, respectively. The vertical black dotted lines indicate the high-velocity (HV, blue/red) and low-velocity (LV, cyan/magenta) ranges, which trace the jet and the outflow, respectively. The definition of LV and HV ranges is based on the SiO ($5-4$) emission (see text).
    {\it Right panels:} Maps of CO ($2-1$), integrated on the LV (top panel) and HV (bottom panel) ranges. First contours and steps are 5$\sigma$, corresponding to 1 Jy km s$^{-1}$ beam$^{-1}$. The black stars indicate the positions of the protostars L1448-C (at the center) and L1448-CS. The black solid line shows the jet/outflow PA \citep{Podio2021}. The beam size is in the bottom-left corner.}
    \label{fig:jets}
\end{figure*}
The sources that make up our sample are part of the Continuum And Lines in Young ProtoStellar Objects IRAM-PdBI Large Program (CALYPSO\footnote{https://irfu.cea.fr/Projets/Calypso/Welcome.html}; \citealt{Maury2019}). CALYPSO is a survey of 16 Class 0 sources, located in different star forming regions (d$\leq$450 pc), observed in three spectral setups (centered at $\sim$ 94, 219 and 231 GHz). The observations were carried out with the IRAM Plateu de Bure Interferometer (PdBI). See \citet{Maury2019} for further details. Out of the sixteen sources, nine can be fully characterized for our purposes. Only for these, in fact, a reliable measure of $\beta$ and of the jets/outflows mass loss rates could be performed ($\dot M_{J}$, $\dot M_{OF}$) (Table \ref{tab:1}). Among the sources considered in this study, seven are in binary or multiple systems. We report considerations on their multiplicity in Appendix \ref{app:binary}.

While this is a low number statistics, we note that CALYPSO is unique in its uniformity and is the only survey for which the SiO~(5\textminus4) transition is systematically targeted to detect the high-velocity jets of a sample of protostars (see Section \ref{sec:outflows}). This allows us to perform the jets/outflows analysis as explained in Sect. \ref{sec:outflows} and have deep enough continuum datasets with which \citet{Galametz2019} measured dust optical properties. Finally, we note that the CALYPSO data for the sources we consider here have been self-calibrated as explained in Section 2 of \citet{Maury2019}, and the self-calibrated data has been later used in \citet{Galametz2019} and \citet{Podio2021}, i.e. the works that have measured the dust opacity spectral index $\beta$ and the jets mass loss rates $\dot M_{J}$ that we also consider in this work.

\section{The dust opacity spectral index}
\label{sec:beta}
\begin{figure*}[ht]
    \centering
    \includegraphics[width=\linewidth]{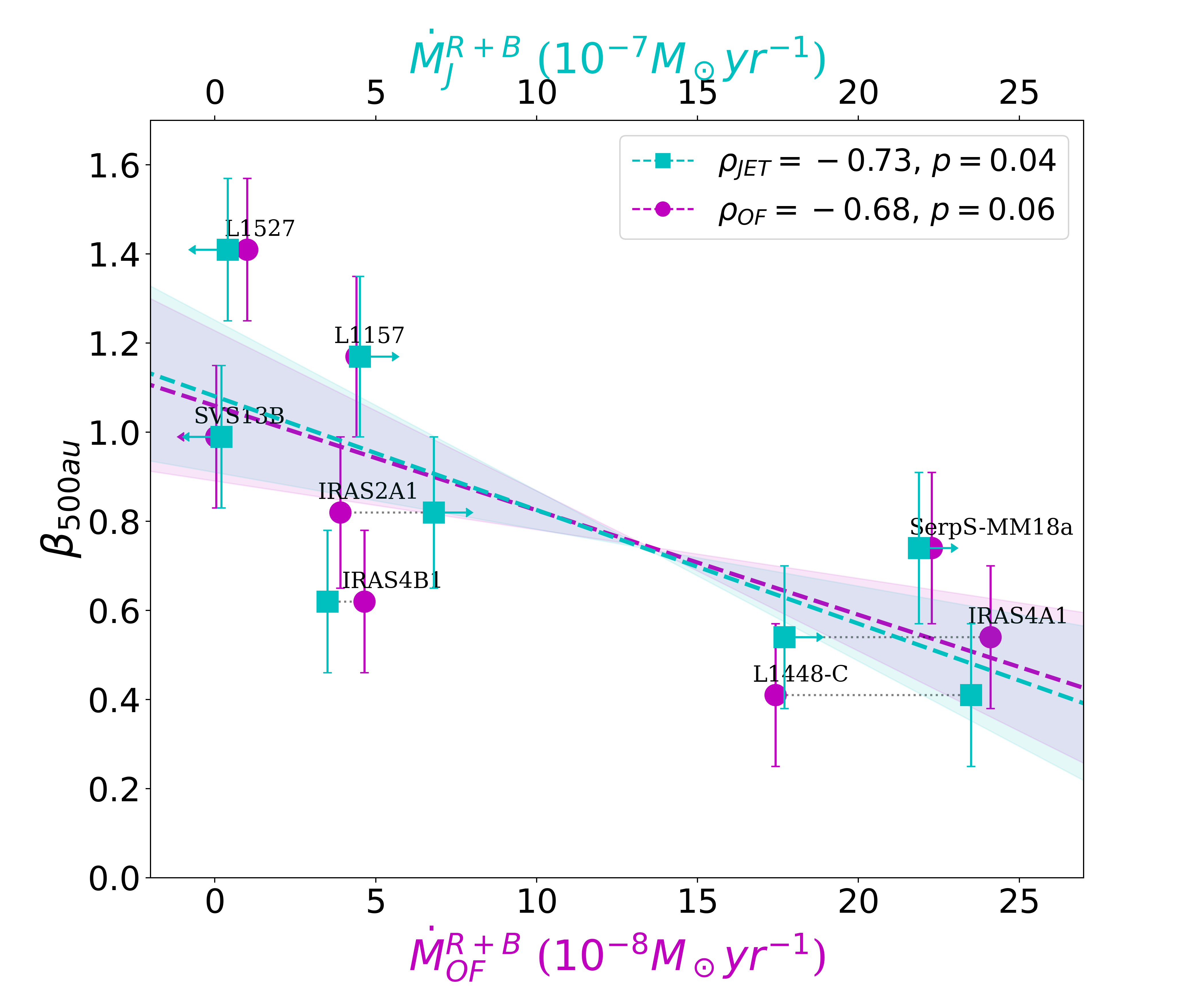}
    \caption{The total (red plus blue lobe) jet mass loss rates (cyan points, upper x-axis) and outflows mass loss rates (magenta points, lower x-axis) around young Class 0 sources anti-correlates with $\beta$ (y-axis) of their inner envelopes ($\beta_{500au}$). For each source, a dotted gray line connects the corresponding jet and outflow rates. Source names are on top of the corresponding magenta point.
    The best-fit linear relations are shown in cyan and magenta, for outflows and jets respectively.}
    \label{fig:correlation}
\end{figure*}
As stated in Section \ref{sec:intro}, the dust opacity spectral index $\beta$ can be derived starting from the radio spectrum of a source and carries dependencies on the properties of interstellar dust, such as the maximum grain size of the distribution. \citet{Galametz2019} used CALYPSO 1.3 and 3.2 millimetre continuum observations to constrain $\beta$ and infer the maximum dust grain sizes in the protostellar envelopes of the sources we consider in this work, up to $\sim$2000 au radial distances from the central protostars. While we report on the details of their measurements in Appendix \ref{app:g19} for completeness, we here briefly summarise them for the reader's convenience.

They measured $\beta$ including a temperature correction that accounts for discrepancies from the Rayleigh-Jeans approximation due to low envelope temperatures:
\begin{equation}
    \beta = \frac{log_{10}\Bigg[\Bigg(F_{\nu_2}/B_{\nu_2}(T)\Bigg) / \Bigg(F_{\nu_1}/B_{\nu_1}(T)\Bigg)\Bigg]}{log_{10}{\nu_2}- log_{10}{\nu_1}}
    \label{eq:beta}
\end{equation}
where $\nu_2=$ 231 GHz and $\nu_1$= 94 GHz are the representative frequencies of the PdBI observations, F$_{\nu}$ is the flux at each frequency. The term B$_{\nu}(T)$ is the value of the Planck function at a temperature T that depends on the radial distance from the central protostar\footnote{In brief, \citet{Galametz2019} assumed this temperature profile based on the radiative transfer post-processing of dusty envelopes from \citet{Motte2001}.} (T $\propto r^{-0.4}$), evaluated at frequency $\nu_i$. For each envelope, they measured $\beta$ across scales and reported best-fit linear models to extrapolate $\beta$ at any other scale.
We report the final estimated envelope-only$\beta$ values at 500 au in Table 1. We note that $\beta$ also depends on the extent of porosity and on the composition of both grain bulk and ice mantles (e.g., \citealt{Natta2007}). However, the lowest values observed by \citet{Galametz2019} are only reconcilable with laboratory experiments in which the sizes of the dust grains are $\gtrsim 100 \mu$m, regardless of ice mantles properties (\citealt{Kholer2015}, \citealt{Ysard2019}). The effect of porosity would also not affect the interpretation of low $\beta$ as due to large grain sizes (e.g., \citealt{Birnstiel2018}).

\section{Jets and Outflows Mass Loss Rates}
\label{sec:outflows}
In this section, we report previous measurements of jets energetics for the CALYPSO sample and present new ones for their low-velocity outflow counterparts. We note that we are interested in the instantaneous mass loss rates of these components rather than their total ejected mass, hence we do not complement the CALYPSO observations with single dish data. This is the case because we aim to investigate a link between the presently observed low $\beta$ values of \citet{Galametz2019} and the continuous flow of material along jets and outflows. The mass loss rates are constant along the jets/outflows extension since mass needs to be conserved, and we measure the latter (outflow) as \citet{Podio2021} measured the former (jet), i.e., at the first peak of the respective tracer emission, to minimize the contribution of possible gas entrained along the jet. The positions of these peaks are in Tab. 4 of Podio et al. (2021). For the reader's convenience, we note that the maximum recoverable scale of the observations is reported to be about 8$\farcs$0 \citep{Podio2021}.
Based on the SiO~(5\textminus4) transition at the innermost knots of the blue- and red-shifted lobes, \citet{Podio2021} defined the high-velocity (HV) ranges, where the emission probes the jet, for all sources associated with SiO (see their Fig. C.1 and Table 4).
In this work, we define the outflow as the emission on the complementary low-velocity (LV) ranges. 
In Fig. \ref{fig:jets}, we show the spatial distribution of CO~(2\textminus1) towards L1448-C, obtained integrating on the LV and HV ranges: HV CO traces the collimated jet, which is believed to originate from the inner disc region, while LV CO probes the wide-angle outflow, which is likely to arise from a more extended disc region. 
The HV ranges for all the CALYPSO sources are listed in Table 4 of \citet{Podio2021}.

At this point, \citet{Podio2021} estimated the $\dot{M_{\rm J}}$, in the blue (B) and red (R) lobes. Here, we apply the same methodology to infer the LV outflows $\dot{M_{\rm OF}}$ of the sources in the CALYPSO sample, for the first time.
The beam-averaged CO column densities in the jet and outflow, $N_{CO}$, are derived from the integrated line intensities on HV and LV, respectively. We assume local thermodynamic equilibrium (LTE) at a fixed excitation temperature, $T_K = 100$~K for HV jets \citep[e.g., ][]{Cabrit2007}, and $T_K = 20$~K for LV outflows \citep[e.g., ][]{Bachiller2001}, and that the emission is optically thin. 
The jet and outflow mass-loss rates are computed as (\citealt{Lee2007}, \citealt{Podio2015}):
\begin{equation}
    \dot M = 1/\sqrt{C} \cdot m_{\unit{H_2}} \cdot (N_{CO}/X_{CO}) \cdot b_t \cdot V_{tan}
    \label{eq:Mjet}
\end{equation}
where  $1/\sqrt{C}$ accounts for  compression in  the shocks ($C=3$), $m_{\unit{H_2}}$ is the mass of molecular hydrogen, 
$X_{CO} = 10^{-4}$ the assumed CO abundance with respect to H$_2$, $b_t$ the beam size perpendicular to the jet, and $V_{tan}$ the tangential jet/outflow velocity. 
The latter is  obtained by correcting for inclination the jet/outflow velocities, assumed to be $100$ km\,s$^{-1}$ for the HV jet, and 10 km\,s$^{-1}$ for the LV outflow.
The inclination is derived from the ratio between the assumed jet velocity and its radial component from the HV spectra (see Table 4 of \citealt{Podio2021}).

For the HV jets, \citet{Podio2021} identified the sources for which $\dot M_{\rm J}$ is a lower limit by comparing CO and SiO spectra. 
The estimated rates carry a factor 3-10 of uncertainty due to the calibration of the parameters of Eq. \ref{eq:Mjet}. The LV outflow emission is likely optically thick, therefore, the estimated $\dot M_{\rm OF}$ must be considered as lower limits. We can estimate the uncertainty introduced by optical depth using $^{13}$CO emission in the assumption that it is optically thin. 
For two sources only (IRAS 4A1 and IRAS 4B1), $^{13}$CO is detected along the jet (see the maps in \citet{Maret2020}).
For these two sources we can reliably estimate the $^{12}$CO/$^{13}$CO ratio, hence opacity. We find $\tau^{(R)}_{IRAS4A1} = 6$, $\tau^{(B)}_{IRAS4A1} = 18$, $\tau^{(R)}_{IRAS4B1} = 15$, $\tau^{(B)}_{IRAS4B1} = 7$. These values imply $\dot M_{OF}$ higher by a factor at least $\sim$6-18. Since we cannot repeat this analysis for all sources, we here stress that the derived $\dot M_{OF}$ (in Table \ref{tab:1} and Fig. \ref{fig:correlation}) are lower limits and we consider the jets to be a more robust proxy of the effective mass loss rates of each protostar. Observations of optically thin tracers of the low-velocity outflows will be key to further test the correlation we propose.

\section{A tentative anti-correlation}
\label{sec:corr} 
Modern theoretical efforts have shown how growing dust grains in protostellar envelopes is problematic due to the lifetimes and densities of these environments (\citealt{Ormel2009}, \citealt{Bate2022}, \citealt{Silsbee2022}, \citealt{Lebreuilly2023}). If millimeter dust, implied by recent measurements of low dust opacity spectral indices in envelopes (\citealt{Miotello2014}, \citealt{Galametz2019}), cannot grow at envelope scales, alternative processes might explain their presence therein. We here show a tentative anti-correlation between $\beta$ with $\dot M_{J}$ and $\dot M_{OF}$, potentially supporting a scenario in which protostars launching powerful outflows can lift millimeter grains into their envelopes.
Fig. \ref{fig:correlation} show the $\beta$ indices found by \citet{Galametz2019} as a function of $\dot M_{J}$ and $\dot M_{OF}$ summed over the blue and red lobes (see Section \ref{sec:sample}). The values are reported in Table \ref{tab:1}. We do not include SerpM-S68N because SiO~($5-4$) emission in \citet{Podio2021} is only at low velocities, likely due to the system inclination, thus impeding the identification of the LV and HV. 
 
The resulting Pearson correlation coefficients are:
\begin{itemize}
    \item[$\bullet$] $\rho_{J}  \hspace{0.2cm} = -0.73 \pm 0.27$ ($\beta_{500au}$, $\dot M_{J}^{R+B}$)
    \item[$\bullet$] $\rho_{OF} = -0.68 \pm 0.28$ ($\beta_{500au}$, $\dot M_{OF}^{R+B}$)
\end{itemize}

We evaluate the statistical significance of such a correlation by means of a two-tailed Student's t-test, where the null hypothesis is that $\rho = 0$ (against $\rho \neq 0$). 
We reject the null hypothesis at $p < 0.04$ level in the jet case, and at the $p < 0.06$ level for the outflows. 
These tentative correlations might support a dust transport scenario from young discs to their embedding envelopes.

Alternative explanations to the observed tentative correlation are possible in case these two share correlations with other quantities. \citet{Bontemps1996} found a correlation between the envelope mass of Class 0/I YSOs and the CO momentum flux of their outflows. Since \citet{Galametz2019} observed a correlation between $\beta$ and envelope mass of CALYPSO sources, then the tentative correlation we show in Fig. \ref{fig:correlation} might be the combined result of these underlying relationships. However, it remains unclear whether the fundamental causal correlation is the one between the dust opacity spectral index and envelope mass or the mass loss reates, as presented here. Moreover, the $\beta-\dot M_{OF}$ correlation in Fig.\ref{fig:correlation} might be caused by an underlying $\dot M_{J}-\dot M_{OF}$ correlation. Such a correlation cannot be quantified here, given that the estimated $\dot M_{OF}$ are lower limits. To rule out possible contamination of the correlation from any dependence of the measured $\beta$ and mass loss rates on the inclination of the source (disc/jet), we reject possible underlying correlations in Appendix \ref{app:inclinations}.

\section{Discussion}
\label{sec:discuss}

We here further discuss our findings, and the conditions that need to be met in order for the proposed dust transport to happen.

\subsection{When and where do transported grains grow?}
\label{sec:timegrow}
If outflows are lifting millimetre (or larger) dust grains into the envelopes of Class 0 objects, these must have first grown in the disc. 
\citet{Brauer2008} studied dust coagulation in the first 1 Myr of disc evolution at representative 1, 10 and 100 au scales and found that millimetre dust grains dominate the dust size distribution already after few $10^3$ yr in the inner 1 au of the disc. \citet{Lebreuilly2023} considered several dust size distributions and simulated their early evolution during protostellar collapse under the effects of turbulent, brownian and radial motions. They found that millimetre grains are formed at $\leq 0.1$ au scales in few years after the first Larson core formation start. Laboratory experiments have been performed to constrain the stickiness of dust grains in the disc inner regions. When heated at 1000 K, dust grains become \textit{super-dry} and their stickiness can increase up to a factor 100, thus providing the conditions to grow even larger agglomerates (\citealt{Bogdan2020}, \citealt{Pillich2021}). 
These temperatures are typically reached in the inner $\sim 0.1$ au of low-mass protostellar discs. At these distances, both jets and outflows could lift grains. Indeed, the typical foot-point of jet is much closer to the star than the outflows'. For example, \citet{Lee2017} measured a $0.05-0.3$ au foot-point radius for the high-velocity SiO jet in the Class 0 HH212 source. Low-velocity outflows, instead, likely extend to a wider disc region out to radii of even $20-40$ au (\citealt{Bjerkeli2016}, \citealt{Tabone2017}, \citealt{Lee2018}), and thus could entrain grains from a larger reservoir.

\subsection{Can outflows lift millimetre grains?}
\citet{Wong2016} and \citet{Giacalone2019} presented an analytical treatment in which they explored the conditions for the uplifting of dust grains along outflows. \citet{Wong2016} presented the critical mass of the protostar for which, if $M_{*} < M_{cr}$, grains of a given size could be entrained against gravity (see their Eq. 7). Another analytical model, by \citet{Giacalone2019}, reported an equation to compute the maximum grain size $a_{max}$ that a given wind can uplift against the gravity of a star of mass M$_*$.
We report the latter for the reader's convenience:
\begin{equation}
    \begin{split}
   a_{max} \approx 0.35 \mu m \Bigg(\frac{M_*}{M_{\odot}}\Bigg)^{-1}  \Bigg(\frac{\dot M}{10^{-8} M_{\odot}/yr} \Bigg) \Bigg(\frac{T_{gas}}{200 K} \Bigg)^{0.5}  \\
    \Bigg(\frac{r}{au} \Bigg)^{-0.25}   \Bigg( \frac{(z/r)}{0.06} \Bigg)^{-1} \Bigg(\frac{\log(r_+/r_-)}{10^3} \Bigg)^{-1},
    \end{split}
    \label{eq:giacalone}
\end{equation}
where $\dot M_{\odot}/yr$ is the mass loss rate of the outflow, $T_{gas}$ is the gas temperature, $r$ is the launching footpoint, $z/r$ the disc flaring ratio, $r_+/r_-$ the ratio between disc's outer and inner edge. See \citet{Giacalone2019} for the details.
We notice that the three sources of our sample with the largest outflows mass loss rates ($\gtrsim$ 2 $\cdot$ 10$^{-7} M_{\odot}/yr$) have $\beta < 0.8$. If we consider this value in Eq. \ref{eq:giacalone}, and we fix M$_* = 1$M$_{\odot}$, T$_{gas} = 20$K at the outflow's base, r $=$ 1 au, z/r = 0.1, $r_+ = 50$ au (typical Class 0 disc radius, e.g., \citealt{Maury2022}) and $r_- = 0.1$ au, we obtain $a_{max} \gtrsim 150 \mu$m. Since outflows mass loss rates are lower limits due to optical depth effects, $a_{max}$ could be higher by even an order of magnitude. We refrain from evaluating Eq. \ref{eq:giacalone} source by source since it was derived for class II objects rather than class 0/I, and because most parameters suffer from large uncertainties for young sources. Thus, at face value, assuming standard parameters, grains larger than 100 $\mu$m could be lifted for the sources with highest mass loss rates (and lowest betas).

Similar findings for the maximum sizes of dust grains entrainable by outflows were reported by \citet{Lebreuilly2020} and \citet{Tsukamoto2021}. They both performed magneto-hydrodynamical simulations. \citet{Lebreuilly2020} ran their setup including large grains to account for growth that might have happened at earlier times, while \citet{Tsukamoto2021} models dust coagulation. They both found that large grains in the inner region of disc (a few 100~$\mu$m to 1 cm) can be entrained.
These grains then decouple from the gas and are ejected from the outflow into the envelope, before falling back into the disc like \textit{ash fall}, as coined by \citet{Tsukamoto2021}.

\subsection{Do grains survive the transport?}
Given their lower velocities and temperatures, as well as a wider entraining base, outflows seem to be the preferred mechanism to lift dust grains from protoplanetary discs to the inner envelopes of young protostars (\citealt{Wong2016}, \citealt{Lebreuilly2020}, \citealt{Tsukamoto2021}). The tentative $\beta - \dot M_{OF}$ correlation we present in Fig. \ref{fig:correlation} might support this scenario. The observed $\beta - \dot M_{J}$ correlation might either mean that jets are contributing to the mechanism or that they share an underlying correlation with the outflows. We thus here discuss if lifted dust grains would survive the transport along jets. Given the much lower speeds and temperatures of outflows, their survival to transport along the latter is a consequence.

The destruction of silicon-bearing dust grains in shocks has been identified as the mechanism that enriches SiO in the interstellar medium and makes of this molecule a key jet tracer \citep[e.g., ][]{Flower1994,Caselli1997,schilke97}.
However, shock models predict that only a small fraction ($< 10\%$) of grains is destroyed in the mild shocks along jets, with typical velocities of $20-50$~km s$^{-1}$ and pre-shock gas densities of $10^{4}-10^6$ cm$^{-3}$ \citep[e.g., ][]{Gusdorf2008a,Gusdorf2008b,Guillet2011}.
 
In the wide grid of models explored by \citet{Gusdorf2008a}, where the shock velocities range $20$~km s$^{-1}$ $< v_{s} < 50$~km s$^{-1}$ and the pre-shock gas densities are in the interval $10^{4}$ cm$^{-3} < n_{H} < 10^6$ cm$^{-3}$, no more than 5\% of Si is released in the gas phase by sputtering. Taking into account shattering and vaporisation of the grains in grain-grain collisions may enhance the fraction of grains destroyed to $\sim 8\%$ \citep{Guillet2011}.
These shock models reproduce the typical SiO abundances estimated in protostellar shocks which span from a few $10^{-8}$ and a few $10^{-7}$ \citep[e.g., ][]{Gibb2004,Bachiller1997,Tafalla2010}. Recent high angular resolution observations, e.g. in CALYPSO, indicate that SiO may reach abundances $>5 \times 10^{-6}$ in jets, which requires either dust-free jets or the fraction of grains sputtered in shocks being larger than $10\%$ (for [Si/H]$_{\odot} \sim 3.5 \times 10^{-5}$, \citealt{Holweger2001}).

Finally, \citet{Wong2016} studied whether (sub-)millimetre dust seeds would survive grain-grain collisions in the envelope, after reaching the transport limit velocity ($v \sim 0.5$ km/s), given by the gravity-drag equilibrium along the outflow. Making use of the shattering model of \citet{Kobayashi2010}, they concluded that millimetre-sized dust grains could survive in the envelope environment: only a fraction as small as 10\% might be destroyed.

Thus, it seems reasonable that a large percentage of dust grains could survive the transport along outflows and even jets, being only partially eroded by collisions with both other grains and gas molecules in the latter. However, we note that there is a strong necessity for dust laboratory and modeling studies to assess the effects of high temperatures in the inner disc if submillimeter dust were lifted from inner outflows footpoints. In particular, it will be crucial to test whether the high temperatures would sublimate grain's mantles materials causing them to further shrink in size.
 
\subsection{Potential implications}
The possibility that protostellar outflows lift large millimetre grains from the disc into the envelopes of young stellar objects can have several implications for the evolution of dust throughout the system.
The outwards transport of dust can extend the timescales of grain growth in discs, limited by the meter barrier problem; it can affect the physical properties of grains as they are transported upwards away from the optically thick disc; and it can contribute to explain mixing of the mineralogy of outer discs, like the one found in meteorites in the Solar System.

First, the orbital dynamics of dust grains orbiting in a disc depends on their stokes number, defined by their composition, density and size. When particles grow in size, they experience a larger and larger headwind that slows them down and cause an inward orbit shift, known as radial drift. In a typical disc orbiting a 1 M$_{\odot}$ star, radial drift velocities of solids at 1 au reach a maximum for meter-sized boulders, thus causing intermediate solids to rapidly fall towards the central star in timescales much faster than the ones estimated for planet cores formation \citep{Weidenschilling1977}. At larger radii, this peak velocity is reached for even smaller pebbles. 
If outflows were uplifting grains in a continuous recycle of dust to the outer disc, this would setback grown millimetre grains in its outskirts and contribute to stretch the available time-span to form larger agglomerates, as already proposed by \citet{Tsukamoto2021}. Moreover, if young protoplanetary discs harbor ring substructures that act like dust traps (as is the case for, e.g., GY91 from \citealt{Sheehan2018} or IRS63 from \citealt{Seguracox2020}), then outwards transported grains will be halted on their drift back towards the inner disc at one of these substructures, potential birthplaces of planetesimals via streaming instabilities \citep{Chambers2021,Carrera2021}.

Secondly, transported dust grains would undergo physical and chemical reprocessing once in the envelope. While they are partially shielded from the radiation of the star in the dense disc, they are going to be lifted in the much thinner envelope and the different energy and intensity of stellar radiation impinging onto them could change their structural and compositional properties. Furthermore, the grains would be transported from the warm inner disc to the colder envelopes where molecular freeze-out could form ice mantles.

Lastly, the uplifting and outward transport of inner disc grains represents a potential explanation for the discovery of cristalline grains in the outskirts of protoplanetary discs, where the temperatures are too low to explain spectral observations of silicate lines (e.g., \citealt{Apai2005}, \citealt{Sargent2009}). Along the same line, \citet{Trinquier2009} and \citet{Williams2020} observed anomalous abundances of $^{46}$Ti, $^{50}$Ti and $^{54}$Cr isotopes in outer Solar System chondrules (mm-sized meteorite inclusions). Since Calcium-Aluminum Inclusions (CAIs), which formed in the inner Solar System, consistently show high abundances in both isotopes, they proposed a mixing of solar nebula material in the early stages of formation. In the same direction are the recent findings of \citet{Hellmann2023}, who show that carbonaceous chondrites display correlation in different isotopes abundances which can be explained by mixing of refractory inclusions, chondrules, and chondrite-like matrix. They thus highlight the need for a mechanism to transport these constituents from the inner disc to its outskirts and trap them in rings where the meteorites would form. If dynamical barriers to outwards viscous transport were present, such as the core of a Jupiter-like planet, protostellar outflows might have played this transport role: the grains extracted by outflows from inner disc regions will later fall back onto the disc out to larger radii.

\section{Conclusions}
\label{sec:conc}
Recently, extremely low dust opacity indices have been observed at few hundred au scales in the envelopes of Class 0 sources, and have been interpreted as the presence of millimetric dust grains \citep{Galametz2019}. Since theoretical models seem to discard the possibility of growing millimetre grains at the densities typical of protostellar envelopes (e.g. \citealt{Ormel2009}, \citealt{Silsbee2022}), we propose here a possible observational test to an alternative explanation, the transport of dust from the disc into envelopes via protostellar outflows. The mechanism has been studied analytically by \citet{Wong2016} and \citet{Giacalone2019} and is supported by numerical simulations of \citet{Lebreuilly2020} and \citet{Tsukamoto2021}. 

We show a tentative anti-correlation between protostellar envelopes $\beta$ and their mass loss rates driven by jets and outflows. Such a correlation can be interpreted as supporting a scenario in which protostellar outflows transport large disc grains into the envelopes of young sources.

If protostellar outflows are indeed lifting millimetre grains in the envelopes of young sources, implications are important for the meter-size barrier problem, the reprocessing of dust during its life cycle, and for material mixing throughout planetary systems, as already suggested for the Solar System (see Sect. \ref{sec:discuss}).
While further measurements of both dust opacity index and mass loss rates will be key in either confirming or disproving such a correlation, we here stress how we explored this possibility with a unique sample in this regard, for which uniform observations, reduction and analyses were carried out. ALMA and JWST synergies will be key to better constrain both dust properties and jets/outflows energetics in a larger sample of young sources.

\section*{Acknowledgments}
This work was partly supported by the Italian Ministero dell’Istruzione, Universit\`{a} e Ricerca through the grant Progetti Premiali 2012-iALMA (CUP C52I13000140001), by the Deutsche Forschungsgemeinschaft (DFG, German Research Foundation) - Ref no. 325594231 FOR 2634/2 TE 1024/2-1, by the DFG Cluster of Excellence Origins (www.origins-cluster.de). This project has received funding from the European Union’s Horizon 2020 research and innovation program under the Marie Sklodowska-Curie grant agreement No 823823 (DUSTBUSTERS) and from the European Research Council (ERC) via the ERC Synergy Grant "ECOGAL" (grant 855130) and from the ERC Starting Grant "MagneticYSOs" (grant 679937).
CC and LP acknowledge the EC H2020 research and innovation
programme for the project "Astro-Chemical Origins” (ACO, No 811312) and the PRIN-MUR 2020 MUR BEYOND-2p (Astrochemistry beyond the second period elements, Prot. 2020AFB3FX).
RSK also acknowledges support from the Heidelberg Cluster of Excellence (EXC 2181 - 390900948) "STRUCTURES", funded by the German Excellence Strategy, and from the German Ministry for Economic Affairs and Climate Action in project ``MAINN'' (funding ID 50OO2206). LC thank Chris Ormel and Sebastian Krijt for insightful discussions on this topic. We thank the referees for their helpful comments, which helped us improved the content and presentation of this work.

%

\vspace{5mm}





\appendix

\section{Binary protostars}
\label{app:binary}
When stars form from the collapse of gas clouds, fragmentation of dense cores often leads to binary or multiple systems. It is estimated that the fraction of stars with at least one companion in the Galaxy is between $\sim$20\% for M-type sources up to $\sim$90\% for O-type ones \citep{Offner2023}. The protostars of the CALYPSO sample are no exception. The PdBI observations beam allowed \citet{Maury2019} to separate systems in the maps with separations larger than $\sim$60 au in Taurus, $\sim$90 au in Perseus and 132 au in Serpens South. For Serpens Main, the systems (SerpM-SMM4, SerpM-S68N) are probed down to distances smaller than 160 au. 
These spatial resolutions are based on distance measurements from \citet{Zucker2019} for Taurus (140 pc), \citet{Ortiz-Leon2018a} for Perseus (290 pc), \citet{Ortiz-Leon2023} for Serpens South (441 pc) and \citet{Ortiz-Leon2018b} for Serpens Main (436 pc).
Moreover, on the large-scale end, they were sensitive to companions up to $\sim$1500-2800 au, depending on the region. They finally classified IRAM04191, L1521F, L1527, L1157, GF9-2, SerpS-MM22 as single sources. On the contrary, L1448-2A, L1448-N, L1448-C, IRAS4A, IRAS4B, SerpS-MM18, SVS13B, IRAS2A were classified as having a companion. 
For each protostar considered in this work, we report the distance of their comapnion(s), if any, in Table \ref{tab:binary}. We note that the tightest binary systems have not been considered here since either a measurement of $\beta$, $\dot M_{J}$, or $\dot M_{OF}$ was impractical in the studies of \citet{Galametz2019}, \citet{Podio2021} or our own, respectively.
While most sources of this study are well resolved binaries, their separation is usually closer than the extent of their envelopes, thus they share a common envelope. The only exceptions are SVS13B and IRAS4B1, for which the companion(s) have much wider separations.
For all the sources considered in this work, and that enter the tentative correlation described in section \ref{sec:corr}, the source of jets and outflows was well resolved (e.g., see Fig. \ref{fig:jets}) and the measurement of $\beta$ could be performed after model subtraction of the secondaries. Furthermore, we note that the low $\beta$ of \citet{Galametz2019} are measured in the inner envelope of each protostar and thus far from possible contamination of the much larger common envelope (see section \ref{sec:beta}).

\begin{deluxetable}{cc}
\tablehead{Source  & Companion(s) name: distance}
\startdata 
L1527      & - \\
L1157      & -  \\
SVS13B     &  SVS13A: 3500au, SVS13C: 4500au\\
IRAS2A1    & IRAS2-A2: 143au \\
SerpS-MM18a& SerpS-MM18b: 2600au$^{(a)}$ \\
SerpM-S68Na & SerpM-S68Nb: 5000au, SerpM-S68Nc: 8300au\\
IRAS4-B1   & IRAS4-B2: 3500au\\
IRAS4-A1   & IRAS4-A2: 420au \\
L1448-C   & L1448-C(S): 2000au\\
\enddata
\caption{Stellar companions associated to the protostars considered in this work. The separations are reported in \citet{Maury2019}. (a): Note that the physical separation of the SerpS-MM18 reported therein should instead be 4420 au, given the most up-to-date distance measurements of the Serpens South region \citealt{Ortiz-Leon2023}.}
\label{tab:binary}
\end{deluxetable}

\section{Details on $\beta$ measurements of Galametz et al. (2019)}
\label{app:g19}
\citet{Galametz2019} measured the dust opacity spectral index in a sample of Class 0/I protostellar envelopes. 
First, for $\beta$ to be a trustworthy proxy of the maximum grain size of a dust distribution, the emission over which the radio spectrum is sampled needs to be optically thin. Hence, they estimated the envelopes optical depths and found $\tau$ well below 0.1 at few hundred au scales for every source (see their Fig. 2).
To make sure the measured $\beta$ would be representative of the envelope alone, \citet{Galametz2019} subtracted both the emission of binary companions (see Section \ref{app:binary}) and circumstellar discs. The companions were subtracted by fitting and removing a gaussian model centered on the secondary sources from the visibilities (further details in \citet{Maury2019}). Secondly, the contribution of the circumstellar disc orbiting the target protostar was evaluated in the uv space as the mean of the amplitudes after 200 k$\lambda$ and subtracted from the shorter-baseline visibilities. They test and comment on the robustness of such a correction in their Section 4, where they assess that considering the mean of the amplitudes in slightly different ranges of the long-baseline-end of the visibilities would not affect their results. Moreover, they subtracted the non-thermal dust contribution by extrapolating literature centimetre data for each source, as shown in their Table A.1.

\section{Jets high velocity ranges}
In Table \ref{tab:ranges}, we report the velocity ranges in which we identify high and low velocity SiO line emission. These ranges were then used to derive mass loss rates from the CO line emission.

\begin{deluxetable}{cccc}
\tablehead{Source  & Lobe & HV range (km/s) & V$_{sys}$ (km/s)}
\startdata 
L1527       & B    & -21.8/-11.8 7 & +5.7\\
            & R    & +17/+27      &\\
L1157       & B    & -60/-20 & +2.6   \\
            & R    & +30/+70  &   \\
SVS13B      & B    & -37/+8.5  & +8.5 \\
            & R    & +8.5/+58  &  \\
IRAS2A1     & B    & -32/-9    & +7.5 \\
            & R    &    -  &      \\
SerpS-MM18a & B    & -17/-2 &  +8.1   \\
            & R    & +21/+32 &    \\
SerpM-S68N  & B    & -7/+5    & +9.2  \\
            & R    & +12/+21   &  \\
IRAS4B1     & B    & -30/-5    & +6.8 \\
            & R    & +16/+50 &    \\
IRAS4A1     & B    & -30/-10  & +6.3  \\
            & R    & +30/+70   &  \\
L1448-C     & B    & -70/-22   & +5.1  \\
            & R    & +25/+85   & 
\enddata
\caption{Table of identified high-velocity ranges from the SiO jet emission for each source (from Tab.4 in \citet{Podio2021}). Based on these ranges, the high- and low-velocity emission of the CO is defined in order to derive the mass loss rates. Only the blue lobe is detected for IRAS2A1. For L1527, no SiO is detected. Therefore the LV range is defined based on the CO emission, while the HV range is assumed to extend +/-10 km/s with respect to the largest velocity detect in the LV.}
\label{tab:ranges}
\end{deluxetable}

\section{Inclination dependencies}
\label{app:inclinations}
As a sanity check for our correlations, we tested whether any underlying correlation exists between the reported ($\beta$, $\dot M_{J}$) or measured quantities ($\dot M_{OF}$) and the inclination of the sources of our sample. To derive the mass loss rates of outflows and jets, in fact, this work and \citet{Podio2021} assumed velocities of 10-100 km/s, respectively (see section \ref{sec:outflows}). \begin{figure*}[t]
    \centering
    \includegraphics[width=\textwidth]{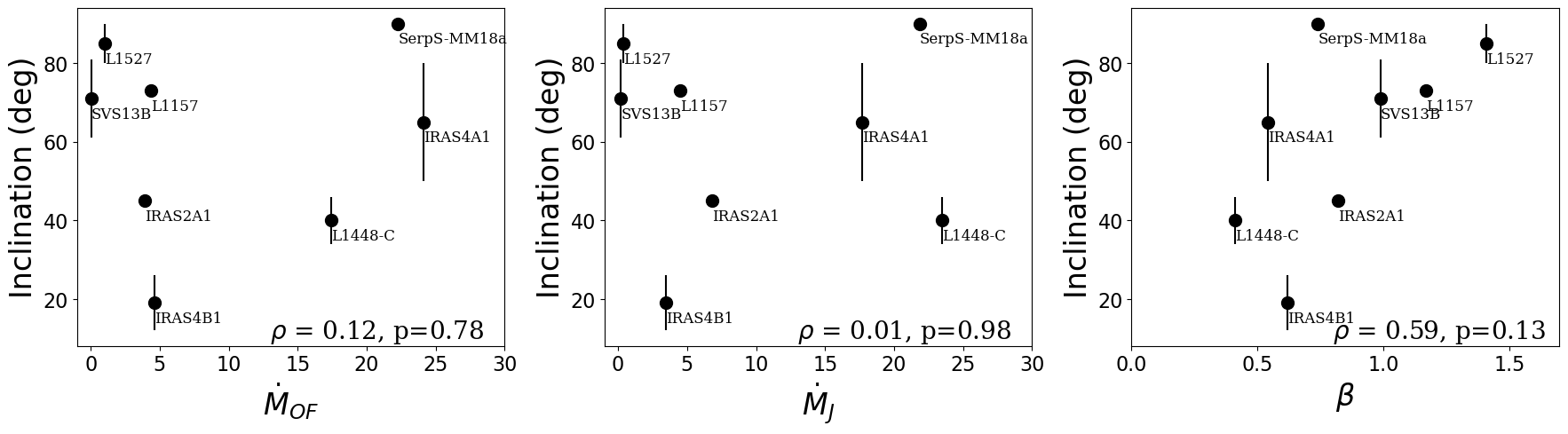}
    \caption{Scatter plot of the jets mass loss rates for the CALYPSO sample with inclinations of their jets (left), and same for the outflows mass loss rates (center) and dust opacity spectral index (right).  The Pearson correlation coefficient and related p-value are reported in the lower right of each panel.}
    \label{fig:corr_test}
\end{figure*}
Such an approach ensures a uniform method to derive the rates, rather than making assumptions on the inclinations of the jets and outflows since current estimates are either unavailable or very uncertain.
Thus, in this section, we check for potential correlation between the quantities involved in our proposed correlation. The inclinations we use have been collected from a number of works. Where more than one estimate was available based on different methods, we reported an average value. If no uncertainty was reported in the literature, for example because the estimate is only of qualitative nature (e.g., for IRAS2A1 reported by \citet{Codella2014}), we plot no error bar. Inclinations for IRAS4A and IRAS4B were reported by \citet{Yildiz2012} and \citet{Marvel2008}. The inclination for the jet of SVS13B was reported to be $\sim$71$^\circ$ by \citet{Segura-Cox2016}, while \citet{podio2016} measured the one for L1157 at $\sim$73$^\circ$. \citet{Yoshida2021} constrained the inclination of L1448-C to be $\sim$34 and $\sim$46 for the blue- and red-shifted lobes, respectively (in this case, we report the main value and scatter between the two). The inclination of L1527 is well constrained to be almost perpendicular to the sky plane (e.g., 85$^\circ$ in \citealt{Ohashi1997}). Finally, \citet{Plunkett2015} and \citet{Podio2021} independently and qualitatively assesed that the jet of SerpM-SMM18a lays in the plane of the sky, so we set it $i =$ 90$^\circ$. Figure \ref{fig:corr_test} summarises the relationship between the inclinations and $\dot M_{J}$, $\dot M_{OF}$, or $\beta$ for the CALYPSO sample.
While a hint for a correlation is seen for the ($\beta$, inclination) pair, only a combination of underlying correlations for both $\beta$ and mass loss rates with inclination would justify the correlation between $\beta$ and mass loss rates. We thus conclude that possible inclination biases are not driving the correlation in Fig.\ref{fig:correlation}.



\bibliographystyle{aasjournal}
\bibliography{biblio.bib}{}

\end{document}